\documentclass[aps,reprint,prl,showpacs,superscriptaddress]{revtex4-2}
\usepackage{amsmath}
\usepackage{amssymb}
\usepackage{graphicx}
\usepackage{color}
\usepackage{bm}
\usepackage{siunitx}
\usepackage{mathtools}
\usepackage{MnSymbol}
\usepackage{cancel}
\usepackage{dcolumn} 
\usepackage{changes}
\usepackage{babel}

\definecolor{linkcolor}{rgb}{0.56640625,0.3359375,0.42578125} 
\usepackage[ pdftex,colorlinks=true,
pdfstartview=FitV,
linkcolor= linkcolor,
citecolor= linkcolor,
urlcolor= linkcolor,
hyperindex=true,
hyperfigures=false]
{hyperref} 

\newcommand{\SRP}{SR-Picard}
\newcommand{\SRPNC}{FES$^{\pm}$}

\begin{document}

\title{Yielding is an absorbing phase transition with vanishing critical fluctuations}

\author{Tristan Jocteur}
\author{Shana Figueiredo}
\author{Kirsten Martens}
\author{Eric Bertin}
\author{Romain Mari}
\affiliation{Univ.~Grenoble-Alpes, CNRS, LIPhy, 38000 Grenoble, France}

\begin{abstract}
   The yielding transition in athermal complex fluids can be interpreted as an absorbing phase transition between an elastic, absorbing state with high mesoscopic degeneracy and a flowing, active state.
   We characterize quantitatively this phase transition in an elastoplastic model under fixed applied shear stress, using a finite-size scaling analysis.
   We find vanishing critical fluctuations of the order parameter (i.e., the shear rate), and relate this property to the convex character of the phase transition ($\beta >1$). 
   We locate yielding within a family of models akin to fixed-energy sandpile (FES) models, only with long-range redistribution kernels with zero-modes that result from mechanical equilibrium. 
   For redistribution kernels with sufficiently fast decay, this family of models belong to a short-range universality class distinct from the Conserved Directed Percolation class of usual FES, which is induced by zero modes.
\end{abstract}

\maketitle

Yield stress fluids make a broad class of soft materials, including emulsions, foams and gels~\cite{coussotYieldStressFluid2014,bonn_yield_2017}. 
Under imposed mechanical stress $\Sigma$, they behave like visco-elastic solids below a yield value $\Sigma_\mathrm{y}$, and otherwise flow like visco-elastic fluids
~\cite{princenRheologyFoamsHighly1989,masonYieldingFlowMonodisperse1996,robertsNewMeasurementsFlowcurves2001}.
Flow 
occurs 
via irreversible plastic events localized in space and time which redistribute stress through elastic interactions~\cite{falk_dynamics_1998,maloneyAmorphousSystemsAthermal2006,tanguyPlasticResponse2D2006,schallStructuralRearrangementsThat2007,lernerScalingTheorySteadystate2009}.
Within the elasto-plastic picture, the yielding transition under imposed shear stress $\Sigma$ (as opposed to imposed shear rate) is an absorbing phase transition (APT)~\cite{nicolas_deformation_2018},
that is, a continuous transition from fluctuating, active states into so-called absorbing
states that trap the dynamics.
In the case of yielding, the active phase is for stresses exceeding $\Sigma_\mathrm{y}$, and the plastic activity $A$, which is proportional to the shear rate $\dot\gamma$, scales as $A \sim (\Sigma - \Sigma_\mathrm{y})^\beta$.
Below the yield value, 
plastic flow stops after a finite strain ($A=0$) 
and the dynamics is frozen in a jammed (absorbing) state.
Elasto-plastic models~\cite{nicolas_deformation_2018} built on this mesoscopic phenomenology 
are minimal models for the statistical physics of yielding, capturing quantitatively plasticity avalanche statistics~\cite{talamali2011avalanches, lin_criticality_2015, liuDrivingRateDependence2016, budrikis2017universal, ferrero2019criticality} and qualitatively the rheology~\cite{picardSlowFlowsYield2005,martensSpontaneousFormationPermanent2012a,nicolasRheologyAthermalAmorphous2014}.

APTs form a broad group of non-equilibrium transitions arising in various areas, e.g., epidemics or fracture propagation ~\cite{marroNonequilibriumPhaseTransitions1999,hinrichsenNonequilibriumPhaseTransitions2006,henkelAbsorbingPhaseTransitions2008}, and spark a continued interest for more than thirty years (recent focus points include hyperuniformity close to APTs~\cite{hexnerHyperuniformityCriticalAbsorbing2015,tjhungHyperuniformDensityFluctuations2015,leiHydrodynamicsRandomorganizingHyperuniform2019a,maTheoryHyperuniformityAbsorbing2023} or the relation to reversible-irreversible transitions in particulate systems under oscillatory driving~\cite{corteRandomOrganizationPeriodically2008,menon2009universality,jeanneretGeometricallyProtectedReversibility2014,mariAbsorbingPhaseTransitions2022,maireInterplayAbsorbingPhase2024}).
A few known universality classes encompass most APTs~\cite{lubeckUniversalScalingBehavior2004,henkelAbsorbingPhaseTransitions2008}.
APTs with short-range interactions, a conserved quantity, and infinitely many absorbing states not related by any symmetry are conjectured to form a universality class called Conserved Directed Percolation (CDP)~\cite{hinrichsenNonequilibriumCriticalPhenomena2000,mannaTwostateModelSelforganized1991,vespignani1998driving,rossi2000universality,le2015exact},
whose exponents are well characterized. 
In particular, the depinning transition of driven elastic manifolds in random media 
~\cite{lin_scaling_2014,lin_criticality_2015,tyukodiDepinningTransitionPlastic2016,ferreroElasticInterfacesDisordered2019},
which shares much of its phenomenology with yielding, 
belongs to CDP~\cite{alavaQuenchedNoiseOveractive2001,bonachelaAbsorbingStatesElastic2007,le2015exact}.

By contrast, it is unclear where yielding stands within the APT landscape. 
Strikingly, it is a convex transition ($\beta>1$) at odds with many other APTs.
Qualitatively, two features distinguish yielding from depinning, both tied to mechanical equilibrium:
the Eshelby kernel exhibits a long-range decay (as $1/r^d$, where $d$ is space dimension), as well as
zero modes~\cite{linDensityShearTransformations2014,lin_scaling_2014,tyukodiDepinningTransitionPlastic2016,ferreroElasticInterfacesDisordered2019}.
Yet, how these features set yielding apart from other APTs, and specifically from those in the CDP class (such as depinning), remains an open issue.

In this Letter, we characterize yielding in $d=2$ as an APT, and show how mechanical equilibrium takes yielding
apart from the CDP class.
Applying a finite-size scaling analysis specifically designed for APTs~\cite{lubeckUniversalFinitesizeScaling2003} to the Picard elasto-plastic model~\cite{picardSlowFlowsYield2005}, we determine the static critical exponents of yielding.
We show that shear-rate fluctuations vanish at the transition, in stark contrast with the CDP class where critical fluctuations diverge.
We then construct a path from yielding to CDP, using as stepping stones variants of the Picard model with non-Eshelby kernels violating the condition of mechanical equilibrium under linear elasticity. 
Making stress redistribution short-ranged or decaying to zero at least as fast as $1/r^6$ leads to an APT class with a concave transition ($\beta<1$), which differs from CDP. 
The latter is only recovered if the redistribution kernel does not have zero modes.
We show how the zero modes and the long-range nature of the stress redistribution kernel modify the stress dynamics (akin to the conserved field dynamics in CDP), paving the way toward a field theory for yielding.

For our demonstration we choose the simplest two-dimensional dynamical model featuring a realistic elastic interaction kernel \cite{tanguyPlasticResponse2D2006,nicolas_deformation_2018}, known as Picard model~\cite{picardSlowFlowsYield2005}. Space is discretized in $N=L^2$ sites arranged on a square lattice $\{\bm{r}_1, \dots, \bm{r}_N\}$ of extension $L$. Sites represent material elements at a mesoscopic scale where continuum quantities are well defined but fluctuating. A site at position $(i,j)$ carries a shear stress $\sigma_{i,j}$, a plastic strain $\epsilon_{i,j}$, and a mechanical state $n_{i,j} = 1$ if  plastic or $n_{i,j} = 0$ if elastic.
The dynamics reads
\begin{align}
    \partial_t\sigma_{i,j} & =  \mu \sum_{k,l} G^\mathrm{E}_{i-k, j-l}\, \dot{\epsilon}_{k,l}\, , 
    \qquad \dot{\epsilon}_{i,j} = \frac{n_{i,j}\sigma_{i,j}}{\mu \tau}\, ,
    \label{eq:picard_model_1}
\\
    n_{i,j}: & \begin{cases}
            0 \xrightarrow{\tau^{-1}} 1 \,, & \text{if } \sigma_{i,j}>\sigma_\mathrm{c} \\
            1 \xrightarrow{\tau^{-1}} 0 \,, & \forall \sigma_{i,j}
            \end{cases} 
\label{eq:picard_model_3}
\end{align}
with $\mu=1$ the elastic modulus, $\tau=1$ the elastic relaxation time, and $\sigma_\mathrm{c}$ a site-independent local yield threshold.
When a site yields, its stress is redistributed according to the discretized Eshelby kernel~\cite{eshelbyDeterminationElasticField1957,picardElasticConsequencesSingle2004}, defined via its Fourier transform $\tilde{G}^\mathrm{E}_{q_x, q_y} =  -4 q_x^2 q_y^2/|\bm{q}|^4$ for $\bm{q} \equiv (q_x, q_y) \neq (0, 0)$. 
Mechanical equilibrium in a continuum material under constant stress
is enforced in discretized and scalar elasto-plastic models by requiring that the sum of the propagator over one of its arguments vanishes, $\sum_{k} G^\mathrm{E}_{k, l} = \sum_{l} G^\mathrm{E}_{k, l} = 0$, which is satisfied here as $\tilde{G}^\mathrm{E}(q_x, 0) =  \tilde{G}^\mathrm{E}(0, q_y) = 0$. 
These constraints are zero modes: any configuration for which the plastic rate is restricted on a line (or column) $m$ on which it is uniform, that is $\dot{\epsilon}_{i,j} \propto \delta_{j, m}$, has no effect on stress values, $\partial_t\sigma_{i,j} = 0$, $\forall i,j$.
Except when specified otherwise, we impose the total stress $\Sigma=\bar{\sigma}_{i,j}$ and measure the shear rate $\dot\gamma = \bar{\dot{\epsilon}}_{i,j}$.

\begin{figure}
    \centering
    \includegraphics[width=\columnwidth]{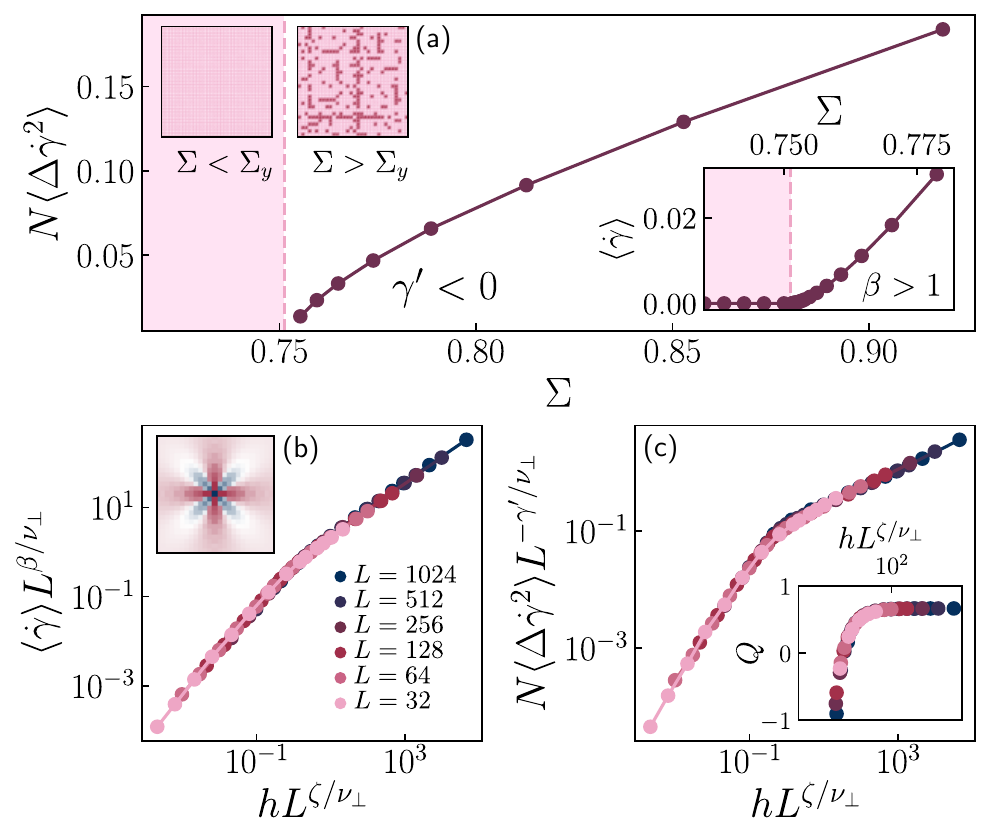}
    \caption{Picard model. (a) Shear rate variance as a function of the imposed stress. In inset: shear rate versus stress. Bottom panels: Finite-size scaling. Scaled shear rate average (b), shear rate variance (c), and cumulant $Q$ (in inset of (c)) as a function of scaled field. In inset of (b): Eshelby propagator in real space. 
}
    \label{fig:raw_Picard}
\end{figure}

We first determine the yield stress in rate-controlled simulations with $L=1024$, where we set $\dot\gamma$, measure $\langle \Sigma \rangle$ and look for the value of $\Sigma_\mathrm{y}$ giving the straightest relation between $\log\dot\gamma$ and $\log\left[\langle\Sigma\rangle - \Sigma_\mathrm{y}\right]$~\cite{SuppMat}.
We then measure the average shear rate $\langle\dot\gamma\rangle$ and shear rate fluctuations, $N\langle \Delta\dot\gamma^2\rangle \equiv N\langle [\dot\gamma - \langle\dot\gamma\rangle]^2\rangle$, in stress-controlled simulations, as shown in Fig.~\ref{fig:raw_Picard}a. 
As expected, we find critical behaviors $\langle\dot\gamma\rangle \sim (\Sigma- \Sigma_\mathrm{y})^\beta$ and $N\langle \Delta\dot\gamma^2\rangle \sim (\Sigma- \Sigma_\mathrm{y})^{-\gamma'}$. 
Surprisingly however, we measure $\gamma' < 0$, that is, rate fluctuations vanish at the transition, in contrast with stress fluctuations under imposed rate, which diverge in the quasi-static limit $\dot\gamma\to 0$~\cite{picardSlowFlowsYield2005, lin_scaling_2014}.
This is an unusual feature of yielding as an APT.

To determine the critical exponents, we perform finite-size scaling, measuring $\langle\dot\gamma\rangle$, $N\langle \Delta\dot\gamma^2\rangle$, and cumulant $Q = 1-\langle\dot{\gamma}^4\rangle/(3\langle\dot{\gamma}^2\rangle^2)$~\cite{lubeckUniversalFinitesizeScaling2003}, varying the value of $L\in [32, 1024]$. 
Because absorbing states prevent determination of steady-state averages close to the critical point, we follow~\cite{lubeckUniversalFinitesizeScaling2003} and introduce a field $h$ allowing for an extra plasticity channel, alongside Eq.~(\ref{eq:picard_model_3})
\begin{equation}
n_i: 0\xrightarrow{h}1,\quad \forall \sigma_i \, .
\end{equation}
This field can be thought of as a result of non-thermally-activated barrier hopping, from e.g., vibration~\cite{goffCriticalityFiniteStrain2019}, activity~\cite{matoz-fernandezNonlinearRheologyModel2017} or local damage~\cite{dallariStochasticAtomicAcceleration2023}, but is here 
merely a process to avoid absorbing states
~\cite{lubeckUniversalFinitesizeScaling2003}.
For $\Sigma=\Sigma_\mathrm{y}$, we expect a critical behavior $\langle \dot\gamma \rangle \sim h^{\beta/\zeta}$.

Under finite-size scaling hypothesis, at $\Sigma=\Sigma_\mathrm{y}$ all curves for different $L$ should collapse  under the rescalings $h \rightarrow h L^{\zeta/\nu_\perp}$, $\langle\dot{\gamma}\rangle \rightarrow \langle\dot{\gamma}\rangle L^{\beta/\nu_\perp}$ and $N\langle \Delta\dot\gamma^2\rangle \rightarrow N\langle \Delta\dot\gamma^2\rangle L^{-\gamma^\prime/\nu_\perp}$.
We then adjust exponents to get the collapse shown in Fig.~\ref{fig:raw_Picard}b--c (see also~\cite{SuppMat} for $Q$). 
We find $\beta\approx 1.5$, in agreement with previous results~\cite{ferreroCriticalityElastoplasticModels2019},  $\nu_\perp \approx 1.14$, 
slightly larger than the value reported for the imposed rate case~\cite{liuDrivingRateDependence2016}, and $\gamma^\prime \approx -0.70$, which to our knowledge was never measured.

These values satisfy the hyperscaling relation $\gamma^\prime = d\nu_\perp - 2\beta$~\cite{fisher1973general,fisherRenormalizationGroupTheory1974,lubeckUniversalScalingBehavior2004}, we find $(2\beta+\gamma^\prime)/(d\nu_\perp) \approx 1.01$. 
Hyperscaling can be rationalized in a scaling scenario assuming that contributors to the global shear rate $\langle \dot\gamma \rangle$ are concurrent but independent quasistatic-like avalanches with spatial extent of order $\xi$~\cite{lin_scaling_2014}.
The correlation length $\xi$ is the distance over which an avalanche induces a significant density of plastic events, beyond $\xi$ the induced plasticity is negligible compared to the average density $\sim \langle \dot\gamma \rangle$.
The duration and period of avalanches of this size scale in the same way, so the local plastic activity, which must remain positive, has fluctuations of the order of its average $\langle \dot\gamma \rangle$, that is, the local variance $\langle\Delta \dot\gamma^2 \rangle_\xi \equiv \langle [\dot\gamma - \langle\dot\gamma\rangle]^2\rangle_\xi \sim \langle\dot\gamma\rangle^2$. 
This leads to the hyperscaling relation via $\langle\Delta \dot\gamma^2 \rangle = (\xi/L)^d \langle\Delta \dot\gamma^2 \rangle_\xi$.
The global variance $N \langle\Delta \dot\gamma^2 \rangle$ thus results from a competition: on a scale $\xi$, the variance of the local plastic activity decreases when approaching the transition, but at the system scale the shear rate results from a decreasing number of independent regions, which tends to increase its variance.
The vanishing of rate fluctuations at the transition is thus due to the fast enough decrease of the variance of local plastic activity as $[\Sigma-\Sigma_\mathrm{y}]^{2\beta}$, and therefore to the large value of $\beta$.
Indeed, vanishing critical fluctuations were reported for few other convex APTs~\cite{argoloVanishingOrderparameterCritical2013,villarroelCriticalYieldingRheology2021a,mariAbsorbingPhaseTransitions2022}.

\begin{figure}
    \centering
    \includegraphics[width=0.5\textwidth]{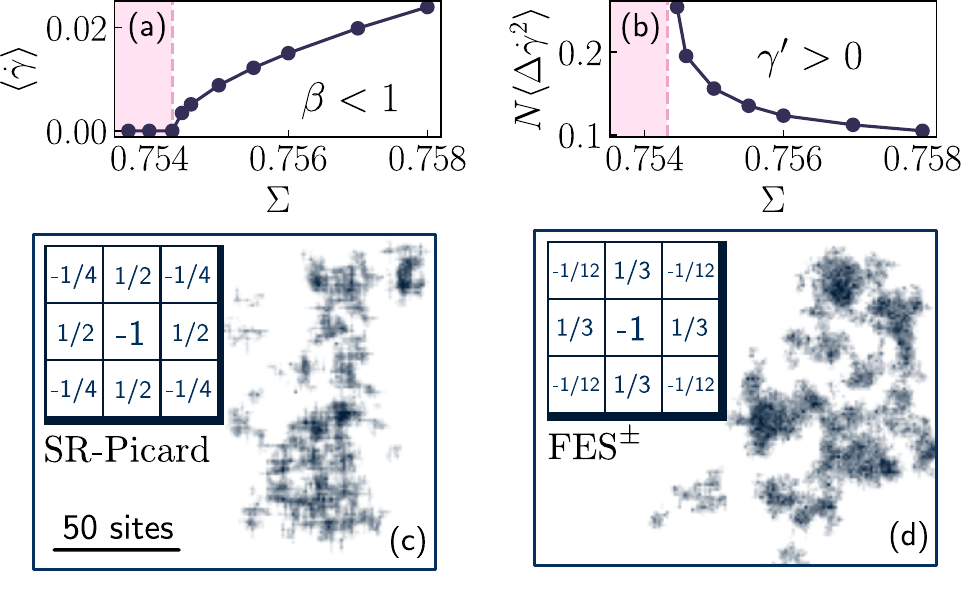}
    \caption{Short-range models. $L=512$ (a) Mean shear rate as a function of  stress for the \SRP{}.  (b) Shear rate variance as a function of stress for the \SRP{}. 
    (c)--(d) activity fields for $h \approx 10^{-9}$ for the \SRP{} (c) and \SRPNC{} (d) models. The activity is averaged over a time span of $100 \tau$. In inset: stress redistribution kernels.
    }
    \label{fig:short_ranged_models}
\end{figure}

To clarify to what extent the specifics of yielding come from the long-range nature of the Eshelby kernel, we turn to a short-range analogue of the Picard model~\cite{martensSpontaneousFormationPermanent2012a} (which we denote \SRP), based on the kernel shown in the inset of Fig.~\ref{fig:short_ranged_models}(c). 
When a site yields, the stress is redistributed only to the eight nearest neighbors, with $G^\mathrm{SR}_{0,0} = -1$, $G^\mathrm{SR}_{0,\pm 1} = G^\mathrm{SR}_{\pm 1, 0} = 1/2$, $G^\mathrm{SR}_{\pm 1,\pm 1} = -1/4$, and $G^\mathrm{SR}_{i, j} = 0$ otherwise.
Crucially, $G^\mathrm{SR}$ retains the quadropular structure of the Eshelby kernel and its zero modes.

The \SRP{} model is superficially similar to fixed-energy sandpile (FES) models~\cite{bakSelforganizedCriticalityExplanation1987,mannaTwostateModelSelforganized1991}, especially those with continuous variables
~\cite{basuFixedEnergySandpilesBelong2012,olamiSelforganizedCriticalityContinuous1992}. 
The stress (mass or energy in FES models) on a plastic site 
is redistributed according to the imposed stress condition, which acts as a conservation law (``fixed-energy''). 
One might thus expect that the \SRP{} falls into the CDP class like FES models~\cite{lubeckUniversalScalingBehavior2004,hinrichsenNonequilibriumPhaseTransitions2006,mannaTwostateModelSelforganized1991,vespignani1998driving,rossi2000universality}.

Performing finite-size scaling~\cite{SuppMat}, we find $\beta \approx 0.59$, $\nu_\perp \approx 0.70$, and $\gamma^\prime \approx 0.26$.
Hence we recover a concave transition ($\beta<1$) with diverging fluctuations ($\gamma'>0$).
Hyperscaling is still satisfied, as $(2\beta+\gamma^\prime)/(d\nu_\perp) \approx 1.03$.
Intriguingly though, these exponent values are not far from, but not quite the CDP values~\cite{lubeckUniversalScalingBehavior2004,henkelAbsorbingPhaseTransitions2008}.
We check that the collapse is significantly worse if we use CDP exponents ~\cite{SuppMat}, which indicates that CDP is not the class of the \SRP~\footnote{The exponents we find for the \SRP{} are close to the ones of the DP class, and indeed the DP exponents provide reasonable (but not the best) collapse~\cite{SuppMat}, but a model with a conserved quantity like \SRP{} is not expected to belong to DP~\cite{henkelAbsorbingPhaseTransitions2008}.}.

\begin{figure}
    \centering
    \includegraphics[width=0.9\columnwidth]{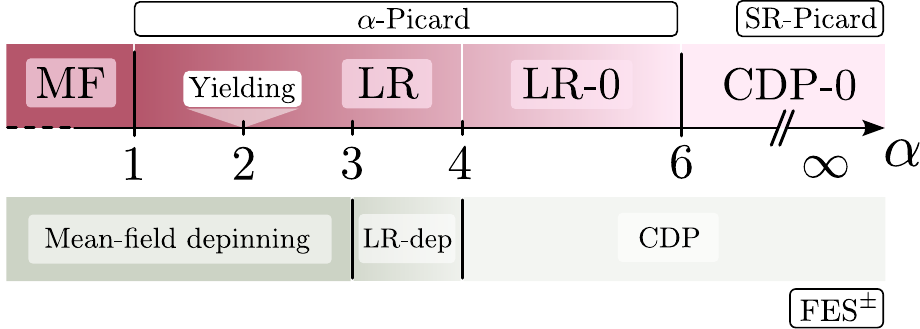}
    \caption{
    Top row: classification of the large-scale behavior of yielding-type models (models with zero modes) as a function of the decay exponent $\alpha$ of the redistribution kernel, summarizing the results of the present work (LR = Long-Range, SR = Short-Range, MF = Mean-Field). Bottom row: similar representation for depinning models (models without zero modes) \cite{caoLocalizationSoftModes2018}. Representative models studied here ($\alpha$-Picard, SR-Picard, FES$^{\pm}$) are also indicated.}
    \label{fig:sketch-landscape}
\end{figure}

A fundamental difference between FES models and stress redistribution models 
is the presence of zero modes of the propagator, 
which consequences are better understood at the continuum level.
In the CDP field theory the conserved field $\rho$ has a non-diffusive dynamics controlled by an activity field $A(\bm{r}, t)$ with $\partial_t \rho = D_\rho\nabla^2 A$, while the activity field follows a Ginzburg-Landau-type dynamics coupled to $\rho$, with multiplicative noise~\cite{vespignani1998driving,van2002universality,rossi2000universality,menon2009universality,le2015exact}.
The Picard, \SRP{} or more generic sandpile models actually share the same stress/mass dynamics on lattice, as given in Eq.~(\ref{eq:picard_model_1}), provided we use the appropriate redistribution kernel $G$
\begin{equation}
  \partial_t \sigma_{i,j} = \sum_{k,l} G_{k,l} A_{i-k,j-l}(t) \, ,\label{eq:naive_stress_dynamics}
\end{equation}
with $A_{i,j}\equiv \dot\epsilon_{i,j}$.
To infer the continuum equations, we introduce a stress field $(a/L)^2\sigma(\bm{r}, t) \equiv \sigma_{i,j}$ and an activity field $(a/L)^2A(\bm{r}, t) \equiv A_{i,j}$,  with $\bm{r}=\{ai/L, aj/L\}$ and $a$ the lattice spacing.
When $G_{k,l}$ is short-ranged
(that is, decays faster than any inverse power law of $s=\sqrt{k^2+l^2}$ for large $s$)
and has no zero modes, a gradient expansion of $A$ shows that in the continuum limit, $L/a\to \infty$, 
we recover the isotropic CDP dynamics for the conserved stress field~\cite{SuppMat}.
By contrast, when $G$ has zero modes, Eq.~(\ref{eq:naive_stress_dynamics}) is dominated by a quartic anisotropic derivative 
\begin{equation}
    \partial_{t} \sigma(\bm{r}, t) = K \partial^2_x\partial^2_y A(\bm{r}, t) \, ,\label{eq:short_range_stress_dynamics}
\end{equation}
with $K = \frac{1}{4}\sum_{k,l} G_{k,l} k^2 l^2$ and with a rescaled time $t \to (a/L)^4 t$~\cite{SuppMat}.
Quantifying how much such modification of the CDP normal form affects the critical behavior  is challenging and out of the scope of this work, but it is reasonable to expect that it does induce a class distinct from CDP, which we call CDP-0 [Fig.~\ref{fig:sketch-landscape}].

\begin{figure}
    \centering
    \includegraphics[width=0.48\textwidth]{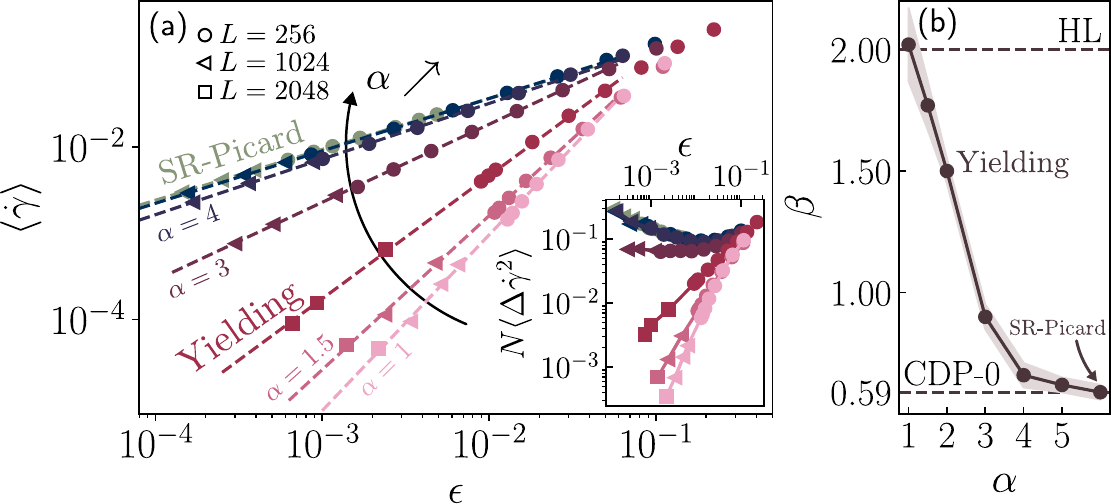}
    \caption{(a) Flow curves $\langle \dot\gamma \rangle$ versus $\varepsilon=(\Sigma-\Sigma_{\mathrm{y}})/\Sigma_{\mathrm{y}}$ for the $\alpha$-Picard model with $\alpha = [5, 4, 3, 2, 1.5, 1]$ and for the \SRP{}. (b) Critical exponent $\beta$ as a function of the decay exponent $\alpha$ of the redistribution kernel.}\label{fig:long-range_exponent}
\end{figure}

To numerically evidence that CDP is recovered without zero modes, we modify the \SRP{} kernel to remove them, as shown in Fig.~\ref{fig:short_ranged_models}(d). The resulting model, that we coin \SRPNC{} to outline the alternate-sign redistribution kernel, unambiguously falls into CDP (see Table~\ref{table:exponents} for exponents values and~\cite{SuppMat} for the finite-size scaling).
Furthermore, as shown in Fig.~\ref{fig:short_ranged_models}(c)--(d), comparing the large scale critical behavior of the \SRP{} and \SRPNC{} models reveals 
the trading of the isotropic CDP dynamics for the anisotropic one in Eq.~(\ref{eq:short_range_stress_dynamics}): While for the \SRPNC{} plasticity localizes in isotropic clusters [Fig.~\ref{fig:short_ranged_models}(d)], for the \SRP{} a distinctive anisotropy along $\bm{e}_x$ and $\bm{e}_y$ is visible
[Fig.~\ref{fig:short_ranged_models}(c)].

\begin{table}
\caption{Measured critical exponents.}
\begin{ruledtabular}
\begin{tabular}{c d d d d} 
  \textrm{Model/Class} & \multicolumn{1}{c}{$\beta$} &
\multicolumn{1}{c}{$\gamma^\prime$}&
\multicolumn{1}{c}{$\nu_\perp$} &
\multicolumn{1}{c}{$\zeta$} \\
 \hline
 Picard/Yielding & 1.5 & -0.70 & 1.1 & 2.5 \\ 
 SR-Picard & 0.59 & 0.26 & 0.70 & 2.1 \\
 FES$^\pm$ & 0.62 & 0.36 & 0.79 & 2.25 \\
 CDP~\cite{henkelAbsorbingPhaseTransitions2008} & 0.64 & 0.37 & 0.80 & 2.23 
 \label{table:exponents}
\end{tabular}
\end{ruledtabular}
\end{table}

Coming back to models with zero modes,
we now argue that yielding belongs to a continuum of classes associated with long-range stress redistribution kernels with algebraic decay, much like equilibrium models with long-range interactions~\cite{fisherCriticalExponentsLongRange1972}. 
We first note that Eq.~(\ref{eq:short_range_stress_dynamics}), derived for a short-range kernel, is also valid for a power-law kernel decaying as $1/s^{\alpha}$, as long as $K$ is finite, that is for $\alpha>6$. To deal with the case $\alpha<6$,
we introduce an $\alpha$-Picard model by considering a kernel $\tilde{G}_\alpha(q_x, q_y) = -b_\alpha q_x^2 q_y^2/q^{6 - \alpha}$ in Fourier space, with $1<\alpha<6$. 
In real space and in the continuum limit $L/a \to \infty$, this corresponds to a kernel $(a/L)^\alpha \mathcal{G}_\alpha(\bm{r}) \equiv G_\alpha(i,j)$  with $\bm{r} = (r\cos\theta, r\sin\theta) \equiv (ai/L, aj/L)$, $\mathcal{G}_\alpha(\bm{r}) \propto [C_\alpha + \cos 4\theta]/r^{\alpha}$ and $C_\alpha$ a constant. 
Yielding corresponds to $\alpha = 2$ (in this case $C_2 = 0$).
The critical behavior for $\langle \dot\gamma \rangle$ in the $\alpha$-Picard model is shown in Fig.~\ref{fig:long-range_exponent}(a) for different values of $\alpha$, and the corresponding exponents $\beta$ are given in Fig.~\ref{fig:long-range_exponent}(b).
Varying $\alpha$, we observe a long-range interaction regime, with $\beta$ being a continuous function of $\alpha$ interpolating from the short-range value $\beta\approx 0.59$ for large $\alpha$ values to $\beta \approx 2$ for $\alpha = 1$, reminiscent of the behavior of avalanches exponents in a statistically isotropic model of yielding~\cite{linDensityShearTransformations2014}.
The value $\beta=2$ is the mean-field value observed for instance in the Hébraud-Lequeux model~\cite{hebraud_mode-coupling_1998}, which suggests that for $\alpha \lesssim 1$ the model is mean-field (range denoted MF in Fig.~\ref{fig:sketch-landscape}).
In parallel to the increase of $\beta$ when $\alpha$ decreases, the fluctuations turn from diverging to vanishing at the transition ($\gamma'$ changes sign) around $\alpha \approx 3$ (inset of Fig.~\ref{fig:long-range_exponent}(a)), which is also roughly where the transition turns convex ($\beta\approx 1$).

Following the reasoning that led us to Eq.~(\ref{eq:short_range_stress_dynamics}), 
the large scale behavior for $\alpha<6$ is of the form~\cite{SuppMat}
\begin{equation}
    \partial_t \sigma = 
    \int \mathrm{d}\bm{s}\, \mathcal{G}_\alpha(\bm{s}) \mathcal{F}[A]\, ,
    \label{eq:long_range_stress_dynamics}
\end{equation}
with a rescaled time $t \to (a/L)^{\alpha-2} t$, and where the functional $\mathcal{F}[A]$ depends on the range of $\alpha$.
For $4 < \alpha < 6$, the presence of zero modes implies
\begin{equation} \label{eq:FA:alpha4:6}
\mathcal{F}[A] = \Delta A(\bm{r}, \bm{s}, t) - \frac{s_\alpha s_\beta}{2}\partial^2_{\alpha\beta} A(\bm{r}, t)
\end{equation}
with $\Delta A(\bm{r}, \bm{s}, t) = A(\bm{r}-\bm{s}, t) - A(\bm{r}, t)$.
This long-range behavior is specific to the zero mode property of the kernel, since when relaxing it, the stress evolution equation takes the CDP form
for $\alpha>4$~\cite{SuppMat}, consistently with depinning with power-law decaying kernels which follows CDP for these values of 
$\alpha$~\cite{tanguyIndividualCollectivePinning1998,caoLocalizationSoftModes2018,lepriolSpatialClusteringDepinning2021}.
This confirms that the presence of zero modes defines for $4<\alpha<6$ a continuum of classes (denoted LR-0 in Fig.~\ref{fig:sketch-landscape}) distinct from depinning/CDP, and characterized by Eqs.~(\ref{eq:long_range_stress_dynamics}) and (\ref{eq:FA:alpha4:6}).
Numerically, we find only a mild dependence of $\beta$ on $\alpha$ in this regime [Fig.~\ref{fig:long-range_exponent}(b)].
In contrast, for $1<\alpha < 4$ (which includes yielding), the dependence of $\beta$ on $\alpha$ is steeper (range denoted LR in Fig.~\ref{fig:sketch-landscape}).
In this regime, we get $\mathcal{F}[A] = \Delta A(\bm{r},t)$, a result that does not rely on the presence of zero modes.
More precisely, a kernel with zero modes leads to the same scaling in $(L/a)^{\alpha-2}$ 
for the convolution in Eq.~(\ref{eq:long_range_stress_dynamics}) as a kernel without zero modes \cite{SuppMat} (in contrast with the case $\alpha>4$ where scalings differ and lead to two distinct classes).
This could mean that depinning and yielding belong to the same class for $\alpha < 4$, but simulations rather support a continuum of classes between pure mean-field depinning and pure yielding ($\alpha=2$) in $d=2$~\cite{ferreroElasticInterfacesDisordered2019}.
For the $\alpha$-Picard model, we find numerically that the long-range behavior for which there is a dependence of $\beta$ on $\alpha$ holds for $1<\alpha<6$ [Fig.~\ref{fig:long-range_exponent}(b)], whereas for depinning models this holds only for $3<\alpha <4$ [LR-dep in Fig.~\ref{fig:sketch-landscape}] and the behavior is mean-field for $\alpha<3$~\cite{caoLocalizationSoftModes2018}.

To sum up, we quantitatively characterize yielding as an APT by performing an extensive finite-size scaling, and find that critical fluctuations vanish at the transition, a property related to the convexity of the transition due to the hyperscaling relation that we verify numerically.
Using both large-scale numerical simulations and an analytical derivation of the continuum stress field dynamics, we provide a classification of yielding-type APTs with zero modes as a function of the stress redistribution range, to locate yielding in a broader APT landscape, as summarized
in Fig.~\ref{fig:sketch-landscape}. We also find that the short-range yielding-like class is distinct from CDP/depinning, pointing to the key role of zero modes for short-range models.

Our results on stress field dynamics constitute a first step towards a field theory for yielding.
Such a field theory would predict stress correlations under slow flows, for which the system is anisotropic, or for yield stress fluids at rest, as they carry anisotropic residual stresses accumulated during past flows~\cite{chungMicroscopicDynamicsRecovery2006,mohan_microscopic_2013,ballauffResidualStressesGlasses2013,vasishtResidualStressAthermal2022}.
Stress correlations in isotropic amorphous materials at rest show a long-range spatial structure mimicking the Eshelby kernel~\cite{lemaitreStructuralRelaxationScaleFree2014,chowdhuryLongRangeStress2016,lemaitreInherentStressCorrelations2017,lemaitreStressCorrelationsGlasses2018,degiuliEdwardsFieldTheory2018,nampoothiriEmergentElasticityAmorphous2020,lernerSimpleArgumentEmergent2020}.
Using Eq.~(\ref{eq:long_range_stress_dynamics}) combined with the linearized CDP activity dynamics as a putative
field theory for yielding, we get a fluctuating hydrodynamics estimate of stress correlations $\langle \tilde{\sigma}(\bm{q})\tilde{\sigma}(-\bm{q})\rangle_\mathrm{c} \propto -\tilde{G}(\bm{q})/(\langle \dot\gamma\rangle + D q^2)$ with $D$ proportional to the plasticity diffusion coefficient~\cite{SuppMat}, so that the Eshelby structure would also dominate at large scale.
Future works should aim at determining a continuum description of the activity dynamics coupled to the conserved stress, which seems particularly challenging, as well as addressing the role of thermal activation to bridge with the ``solid that flow'' picture of supercooled liquids~\cite{maierEmergenceLongRangedStress2017,chackoElastoplasticityMediatesDynamical2021,ozawaElasticityFacilitationDynamic2023}.

\acknowledgments
This project was provided with computer and storage resources by GENCI at
IDRIS thanks to the grant 2023-AD010914551 on the supercomputer
Jean Zay's V100 and A100
partitions.

\nocite{fisherScalingTheoryFiniteSize1972}

\end{document}